\documentclass[%
 aip,
 amsmath,amssymb,
 reprint,%
]{revtex4-2}

\usepackage{graphicx}
\usepackage{dcolumn}
\usepackage{bm}

\usepackage[utf8]{inputenc}
\usepackage[T1]{fontenc}
\usepackage{mathptmx}
\usepackage{etoolbox}

\makeatletter
\def\@email#1#2{%
 \endgroup
 \patchcmd{\titleblock@produce}
  {\frontmatter@RRAPformat}
  {\frontmatter@RRAPformat{\produce@RRAP{*#1\href{mailto:#2}{#2}}}\frontmatter@RRAPformat}
  {}{}
}%
\makeatother
\begin{document}

\preprint{AIP/123-QED}

\title{Thermomechanical model of solar cells}
\author{Tom Markvart}
 \altaffiliation[Also at the  ]{School of Engineering Sciences, University of Southampton, UK }

\affiliation{%
Centre for Advanced Photovoltaics, Faculty of Electrical Engineering, Czech Technical University in Prague 
}%
\email{tm3@soton.ac.uk}
\date{\today}

\begin{abstract}
\textbf{Abstract} The paper considers a model for the solar cell as a mechanical open-cycle thermodynamic engine where chemical potential is produced in an isochoric process corresponding to the thermalization of electron-hole pairs. Expansion of the beam under one-sun illumination and current generation are described as isothermal lost work.  More generally, voltage produced in an open cycle process corresponds to availability, leading to a correction to the Shockley-Queisser detailed balance limit. 
\end{abstract}

\maketitle
\section{Introduction}
Thermodynamics provides an alternative view of photovoltaic conversion, adding fundamental aspects to the traditional electronic model.\footnote{H.A. Muser,  Thermodynamische Behandlung von Elektronenprozessen
in Halbleiter-Randschichten, Zeitschrift f\:ur Physik, 148, 380 (1957).}$^,$\footnote{T. Markvart, Thermodynamics of losses in photovoltaic
conversion, Appl. Phys Lett 91, 064102 (2007).}$^,$\footnote{T. Markvart, Solar cell as a heat engine: Energy-entropy
analysis of photovoltaic conversion. Phys
Status Solidi (a) 205, 2752 (2008).} Perceiving, in the Shockley Queisser detailed balance,\footnote{W. Shockley and H.J Queisser, Detailed balance limit of
efficiency of p-n junction solar cells. J. Appl. Phys. 32, 510 (1961).} photon fluxes as heat, thermodynamics allows the energy – in other words, voltage -  to be determined in terms of photon entropy. 

In this paper, a thermodynamic model of the solar cell is developed by a direct analogy with the classical thermodynamics of mechanical heat engines.
This theoretical angle becomes possible by employing a relationship between the number of photons $N$ in a volume $V$ representing the solar cell which emits photons in a beam with \'etendue $\cal{E}$ and photon flux $\Psi$:\footnote{T. Markvart, The thermodynamics of optical étendue,
J Opt A: Pure Appl. Opt. 10, 015008 (2008).} 
\begin{equation}
\Psi =\frac{c}{4\pi }{\cal{E}}\frac{N}{V}
\label{correspondce}
\end{equation}

Equation (1) establishes a correspondence, between the photon number $N$ and photon flux density $\Psi$ on the one hand, and between the volume $V$ and the \'etendue on the other. In particular, it permits the definition of a volume $v$ of a “thermodynamic photon” in terms of the beam parameters $\cal{E}$ and  $\Psi$:
\begin{equation}
v=\frac{V}{N}=\frac{c}{4\pi }\frac{{\cal{E}}}{\Psi }	
\label{volume}
\end{equation}

The thermodynamic photons that mean here are thermal photons, similar to a photons of blackbody radiation but restricted to the spectral interval which can be absorbed by the solar cell – in other words, radiation with photon energy in excess of the semiconductor bandgap $E_g$ . We shall call this radiation quasi-blackbody radiation. The volume per photon in a solar cell at temperature $T$ emitting with unit emissivity / absorptivity into a full hemisphere (\'etendue $\pi A$, where $A$ is the cell area) is then
\begin{equation}
v=\frac{c}{4{{\Phi }_{{{E}_{g}}}}(T)}
\label{v-Phi}
\end{equation}
\noindent
where 
\begin{equation}
{{\Phi }_{{{E}_{g}}}}(T)=\frac{\Psi}{A}=\frac{2\pi }{{{h}^{3}}{{c}^{2}}}\int\limits_{{{E}_{g}}}^{\infty }{\frac{{{E}^{2}}}{{{e}^{E/{{k}_{B}}T}}-1}dE}
\label{SQ-density}
\end{equation}
\noindent
is the photon flux density.
 
\section{Chemical work from photons}
We start by writing down the equivalent of the first law of thermodynamics for a photon in the form 
\begin{equation}
du=Tds-pdv
\label{first-law}
\end{equation}
where $T$ and $p$ are the temperature and pressure, and $u$ and $s$ are the energy and entropy per photon. Equation~(\ref{first-law}) will be supplemented by an equation for the free energy per photon – the chemical potential $\mu$ :
\begin{equation}
\mu =u-Ts
\label{chempot}
\end{equation}

Equation~(\ref{chempot}) divides photon energy $u$ into a part - the chemical potential - which can be extracted to carry out useful work or generate voltage. The second term on the right hand side -- proportional to the entropy -- is the heat that needs to be exchanged with a reservoir at the same temperature as the source of photons when a photon is emitted or absorbed. 

The chemical potential replaces mechanical work of classical thermodynamics as the key parameter in a thermodynamic treatment of photovoltaics. To represent this different slant, we  replace photon energy $u$ in~(\ref{first-law}) by the chemical potential~(\ref{chempot}):
\begin{equation}
d\mu =-sdT-pdv
\label{dmu-equation}
\end{equation}

\begin{figure}
\centering    
\includegraphics[width=1\linewidth]{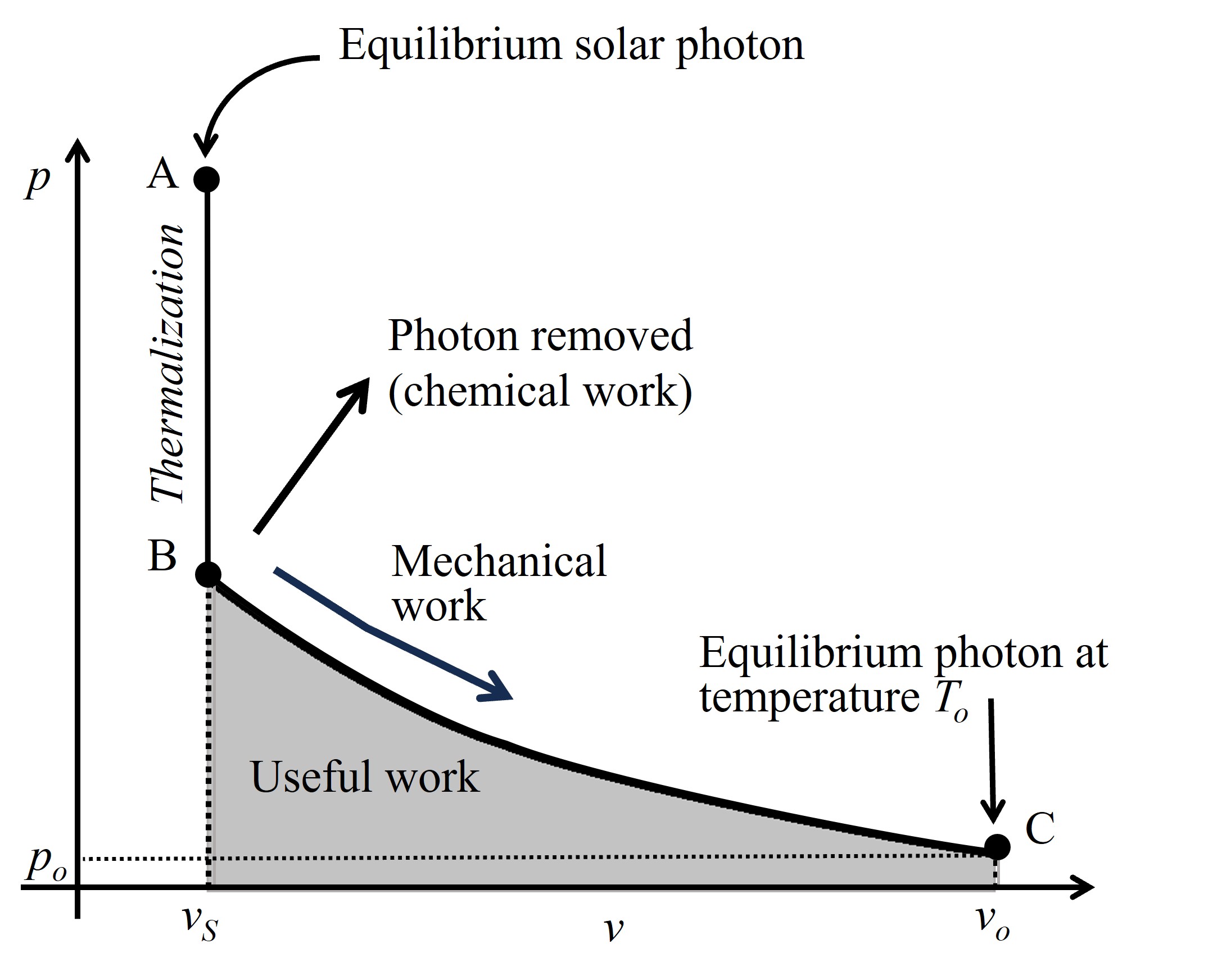}
\caption{Chemical work produced by photon cooling (thermalization) from temperature $T_S$ to temperature $T_o$, compared with mechanical work along the isotherm BC.}
\label{work}
\end{figure}
With the relevant formalism in place, we can now picture a thermodynamic process to mimic photovoltaic conversion, as the conversion of a solar photon at temperature $T_S$ into a photon at ambient temperature $T_o$ whilst producing work. The initial (solar) photon has a chemical potential of zero, corresponding to blackbody radiation. The final  (ambient temperature) photon  has a chemical potential equal to the voltage in energy units at the contact of  the solar cell.

As is customary, we will picture this process  in a $p-v$ diagram for one photon - one photon of quasi-blackbody radiation with an optical bandgap $E_g$.
\footnote{We should emphasize that the $p-v$ diagrams considered in this paper  are not to scale as the actual volume and pressure change by many orders of magnitude.} 
The first part of the conversion process is thermalization of the electron-hole pairs, accompanied by cooling the photon gas to temperature $T_o$. 

We consider here, for the moment, conversion at open circuit when the emitted photon flux is equal to the incident (and fully absorbed) photon flux. Let us also assume that the conversion takes place under maximum concentration of sunlight. This means that the emitted and incident beams cover the full hemisphere, and have the same \'etendue $\pi A$. 

Equation~(\ref{volume}) then tells us that the beam transformation takes place at constant volume, and can be represented by line AB in Fig.~\ref{work}, where the volume $v_S$ is given by Eq.~(\ref{v-Phi}) with $T=T_S$. In mechanical terminology, thermalization is therefore an isochoric process. Constant volume means that no mechanical work is carried out during thermalization. But the photon produces chemical work by increasing its chemical potential from zero at point A to some finite value $\mu_{max}$ at point B. This chemical potential is equal to the chemical potential of electron-hole pairs with which the photons are in equilibrium. 

Following thermalization at point B, the photon gas reaches the final temperature $T_o$ but its density $n^{ph}=1/v_S $ is high. There are now two options: photons can be removed by absorption to produce current.  Alternatively,  the photon gas can perform mechanical work $w$ by isothermal expansion BC to reach complete equilibrium at density $1/v_o$,  ambient temperature $T_o$, and chemical potential zero.
\enlargethispage{1.5\baselineskip}

We can represent the two processes -- thermalization leading to chemical potential  $\mu_{max}$, and the mechanical $w$ work done by expansion BC -- quantitatively by integrating Eq.~(\ref{dmu-equation}).  We observe that each of the two steps AB and BC corresponds integration over one variable $T$ or $v$, giving 
\begin{equation}
{{\mu }_{o}}-{{\mu }_{S}}=0=-\int\limits_{{{T}_{S}}}^{{{T}_{o}}}{sdT}-\int\limits_{{{v}_{S}}}^{{{v}_{o}}}{pdv}
\label{mu-mu}
\end{equation}
since the chemical potential of the equilibrium photons at the beginning and end of the process is zero. 
The second term on the right hand side of~(\ref{mu-mu}) is the mechanical work $w$. The first term gives the chemical potential $\mu_{max}$ produced by thermalization: 
\begin{equation}
{{\mu }_{max}}=\int\limits_{{{T}_{o}}}^{{{T}_{S}}}{s({{v}_{s}},T)dT}
\label{mu-max-eqn}
\end{equation}

Equation~(\ref{mu-mu}) therefore expresses the equivalence of the chemical and mechanical work.
Let us first interpret the meaning of $\mu_{max}$ in the language of photovoltaics. To this end we assume that photon density is sufficiently low and photons form an ideal gas law which, for one photon, reads
\begin{equation}
pv={{k}_{B}}T
\label{ideal-gas}
\end{equation}
Evaluating the second integral in Eq.~(\ref{mu-mu}) then gives
\begin{equation}
{{\mu }_{max}}=w={{k}_{B}}{{T}_{o}}\int\limits_{{{v}_{S}}}^{{{v}_{o}}}{\frac{dv}{v}}={{k}_{B}}{{T}_{o}}\ln \left( \frac{{{v}_{o}}}{{{v}_{S}}} \right)
\label{mu-max-1}
\end{equation}
where the volume $v_o$ is given by~(\ref{v-Phi}) with $T=T_o$. In terms of photon flux densities, 
\begin{equation}
{{\mu }_{max}}=k_B{{T}_{o}}\ln \left\{ \frac{\Phi_{E_g} \left(T_S \right )}{\Phi_{E_g}\left(T_o\right)} \right\}=qV^{max}_{oc}
\label{Vmax-SQ}
\end{equation}
in other words, the Shockley-Queisser open circuit voltage in energy units, under maximum concentration of sunlight.

Equation~(\ref{Vmax-SQ}) was obtained by direct means in \footnote{T. Markvart, Thermodynamic limit on the open-circuit voltage of solar cells, Prog. Photovolt. Res. Appl. 33, 664 (2025).} discussing also its relationship with the Trivich and Flinn\footnote{D. Trivich and P. Flinn,  Maximum efficiency of solar
energy conversion by quantum processes. In:
Daniels F, Duffie J, eds. Solar Energy Research.
London, Thames and Hudson; 1955, 143.} approximation for the voltage, and the voltage generated by hot carrier solar cells. 

\section{Isothermal losses}
We can pursue the mechanical analogy further to model the conversion process more generally,  under one-sun illumination, away from open circuit. At one sun intensity, the solar cell converts an incident beam with \'etendue ${\cal{E}}_{in}=\omega_S\, A$, where $\omega_S$ is the solid angle subtended by the Sun. As the solar cell emits a beam into the full hemisphere (\'etendue ${\cal{E}}_{out}=\pi\, A$), the photon gas expands from the volume $v_S$ under maximum concentration to a volume $v_1=v_S (\pi/\omega_S)$.  

\begin{figure}
    \centering
    \includegraphics[width=1\linewidth]{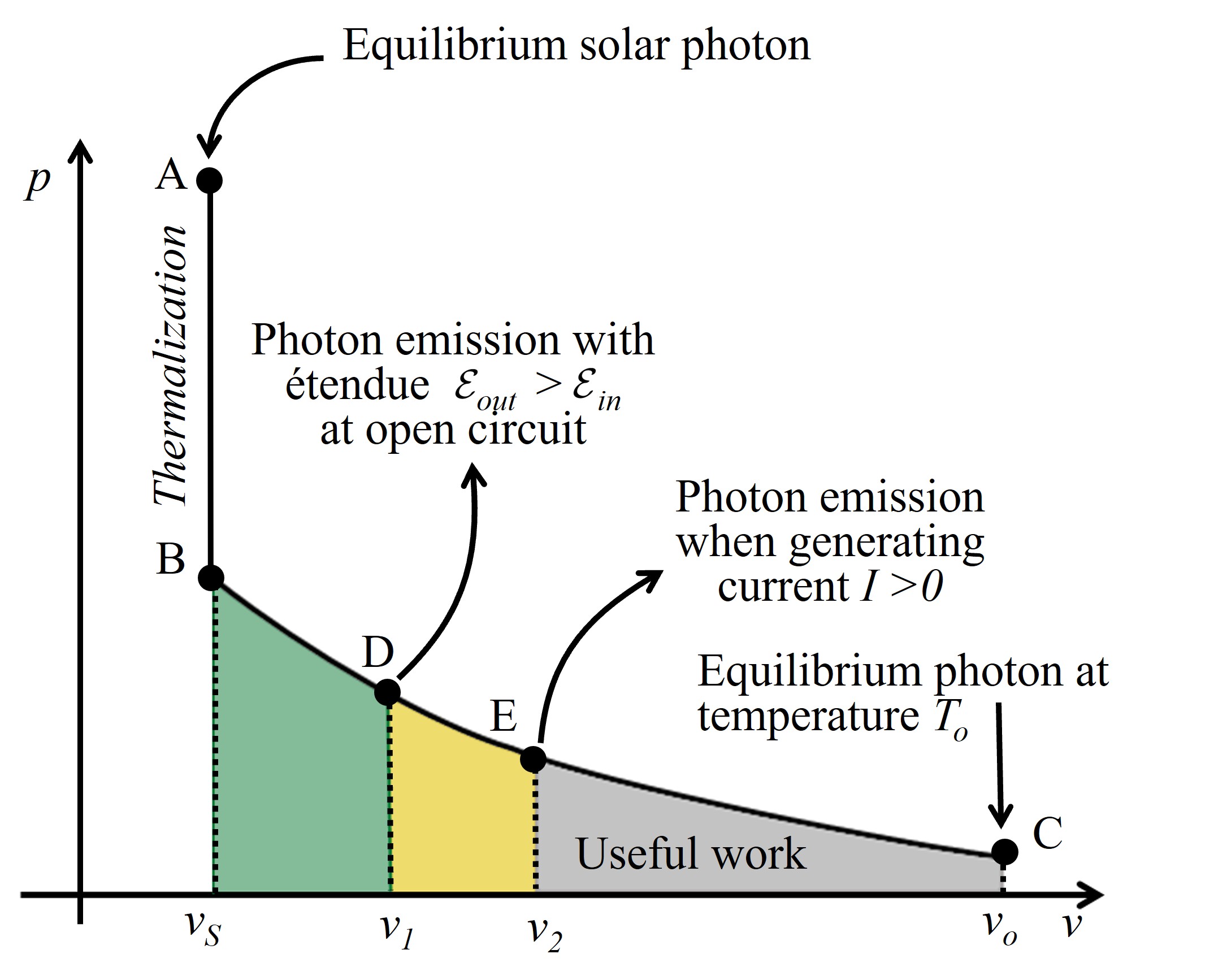}
    \caption{Isothermal losses pictures in the $p-v$ plane.}
    \label{isothermal-losses}
\end{figure}

The chemical potential of the emitted light is again obtained by integrating~(\ref{dmu-equation}), this time along the path ABD in the $p-v$ diagram (Fig.~\ref{isothermal-losses}). The temperature integration produces again the chemical potential $\mu_{max}$ but there is now a further contribution, from the volume integral. This contribution is negative, and represents a "lost work" of magnitude
\begin{equation}
{{w}_{loss}^{(1)}}=\int\limits_{{{v}_{S}}}^{{{v}_{1}}}{pdv}
\label{lost-work-1}
\end{equation}
Using~(\ref{ideal-gas}), the integral~(\ref{lost-work-1}) is easily evaluated to give
\begin{equation}
{{w}_{loss}^{(1)}}={{k}_{B}}{{T}_{o}}\ln \left( \frac{\pi }{{{\omega }_{S}}} \right)
\label{lost-work-12}
\end{equation}
We observe that the conversion of one-sun radiation incurs a loss~(\ref{lost-work-12}) due to the expansion of the light beam in the conversion process. Using our mechanical analogy, this is similar to the lost work in an expansion of gas into vacuum. 

We can now consider current generation. Generation of  current (with density $J$, say) reduces the emitted photon flux density by an amount $J/q$. The emitted photon flux density at point D, $(\omega_S/\pi) \Phi_{Eg}(T_S)$, will therefore be reduced to $(\omega_S/\pi) \Phi_{Eg}(T_S)-J/q$, moving the emission to point E where the volume per photon is increased by
\begin{equation}
\frac{v_2}{v_1}=\frac{(\omega_S/\pi) \Phi_{Eg}(T_S)}{(\omega_S/\pi)\Phi_{Eg}(T_S)-J/q}
\label{lost-work-2}
\end{equation}
with respect to point D. The corresponding lost work is given by
\begin{equation}
{{w}_{loss}^{(2)}}=\int\limits_{{{v}_{1}}}^{{{v}_{2}}}{pdv}=k_BT_o\ln \left( \frac{v_2}{v_1} \right )
\label{lost-work-21}
\end{equation}

We can relate the loss~(\ref{lost-work-21}) to solar cell parameters since
\begin{equation}
\begin{split}
J_{\ell}&=(\omega_S/\pi)\Phi_{Eg}(T_S)-\Phi_{Eg}(T_o)
\\
J_o&=\Phi_{Eg}(T_o)
\end{split}
\label{currents}
\end{equation}
where $J_{\ell}$ and $J_o$ are the photogenerated and dark saturation current densities. The lost work (or loss in chemical potential)~(\ref{lost-work-21}) then becomes
\begin{equation}
{{w}_{loss}^{(2)}}={{k}_{B}}{{T}_{o}}\ln \left(\frac{J_{\ell}+J_o}{J_{\ell}+J_o-J} \right)
\label{lost-work-22}
\end{equation}

Collecting together~(\ref{Vmax-SQ}),~(\ref{lost-work-12}) and~(\ref{lost-work-22}) we obtain
\begin{equation}
\begin{split}
q{{V}_{oc}}=&\,q{V}^{max}_{oc}-{{k}_{B}}{{T}_{o}}\ln \left( \frac{\pi }{{{\omega }_{S}}} \right)
\\
qV=&\,q{{V}_{oc}}-{{k}_{B}}{{T}_{o}}\ln \left( \frac{{{J}_{\ell }}+{{J}_{o}}}{{{J}_{\ell }}+{{J}_{o}}-J} \right) 
\end{split}
\end{equation}
which transforms, using~(\ref{currents}), into
\begin{equation}
J=J_{\ell}-J_o \left(e^{qV/k_BT_o}-1 \right)
\label{lost-work-23}
\end{equation}
in other words, the usual solar cell equation with parameters~(\ref{currents}). In thermodynamic terms, the isothermal losses~(\ref{lost-work-12}) and~(\ref{lost-work-22}) are due to irreversible entropy generation.$^2$  

\section{Discussion}

We have shown  that the thermomechanical model of the solar cell is equivalent to the Shockley-Queisser detailed balance model. This is not surprising as what we have done is to replace the incident and emitted light beams, modelled as quasi-blackbody radiation,  by heat. As the chemical potential of this radiation is zero, the equivalence with heat is evident. 

\begin{figure}[b]
    \centering
    \includegraphics[width=0.9\linewidth]{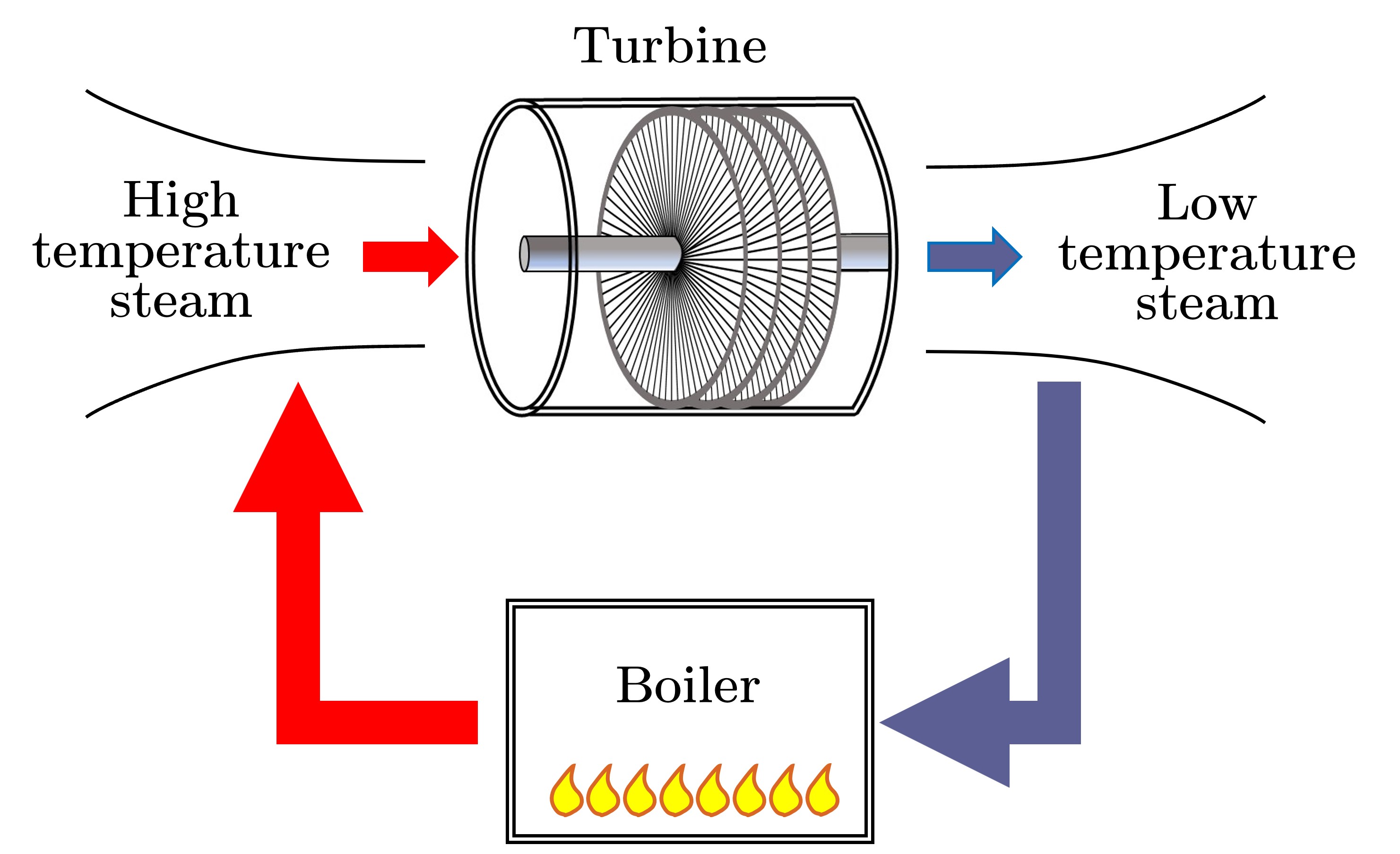}
    \caption{Open and closed cycle conversion, illustrated on the example of a steam turbine. }
    \label{turbine}
\end{figure}
Notwithstanding, it is of interest to compare the present model with heat engines of classical thermodynamics. These engines, in contrast with the present model,  operate in a cyclical manner. Let us consider such a heat engine on the example of a steam turbine (Fig.~\ref{turbine}). A steam turbine converts steam at high temperature and pressure into low temperature steam at the exit from the turbine. The steam then has to be returned to the high temperature inlet of the turbine by heating in a boiler. This closes the cycle and permits a description of the steam turbine as a standard heat engine.  

If  high-temperature steam were to be available free of charge, we can dispense with the boiler and consider only the open cycle consisting of work produced by cooling the high temperature steam to ambient temperature.  The steam -- or, in our case, photons -- acts as both fuel and working medium.\footnote{Interestingly, such open cycle operation is characteristic also for other renewable energy applications such as wind, tidal or geothermal energy.}

A further difference with classical closed cycle engines appears if we examine the medium which surround the engine. In a closed cycle operation, the work carried out by the equilibrium pressure $p_o$ of this medium does not appear by virtue of the fact that the elements of work $p_o\delta V$ cancel in the compression and expansion parts of the cycle. 

This is not the case for an open-cycle converter where the work carried out by the external medium, equal to $p_o(v_o-v_S)$ in Fig.~\ref{work}, should be subtracted from the useful work. This is a well known result of classical thermodynamics which gives the useful work as availability\footnote{See, for example, A.B. Pippard, The Elements of Classical Thermodynamics.
Cambridge,  Cambridge University
Press, 1964.}; it  is also  the reason why the limiting efficiency for the conversion of  full blackbody radiation is the Landsberg efficiency, rather than the Carnot efficiency\footnote{R. Petela, Exergy of heat radiation. Trans. ASME Heat
Transfer, 36,187 (1964)}$^,$\footnote{ P.T. Landsberg and J.R. Mallinson, Thermodynamic constraints,
effective temperatures and solar cells. In: Colloque
International sur l’Electricitè Solaire, CNES,
Toulouse, 1976, 16.} (see also 
\footnote{T. Markvart and G. Bauer, What is the useful energy of a
photon? Appl. Phys. Lett. 101, 193901 (2012)}).   

The "availability correction" to the Shockley-Queisser voltage can easily be estimated to be about $k_BT_o/q\simeq 26$ mV - not large but noticeable, lowering the Shockley-Queisser efficiency limit by about 1\% absolute. 

\section{Acknowledgement}
This work was supported by project "Energy Conversion and Storage", grant no. CZ.02.01.01/00/22\_008/0004617, programme Johannes Amos Commenius, Excellent Research.

\end{document}